\documentclass{PoS}

\usepackage{setspace}

\title{Usage of tracking in the CMS Level-1 trigger for the High Luminosity LHC Upgrade}

\ShortTitle{CMS Level-1 track trigger for the HL-LHC Upgrade}

\author{\speaker{Chang-Seong Moon}\thanks{on behalf of the CMS Collaboration. This research was supported by Kyungpook National University Research Fund, 2017}\\
        Kyungpook National University\\
        E-mail: \email{csmoon@cern.ch}}

\abstract{At the High Luminosity LHC (HL-LHC), the CMS experiment will face a harsh environment with a high instantaneous luminosity up to 7x10$^{34}$/cm$^2$/s corresponding to an average of 140-200 multiple proton-proton collisions per bunch crossing. The main goal of the CMS Level 1 (L1) trigger upgrade for the HL-LHC is to maintain trigger thresholds that are as low as possible and comparable to those currently in use at the LHC, and to possibly include new triggers that were not feasible at the LHC. This will be achieved by upgrading the detector readout electronics, to allow a much larger L1 trigger rate, and by including, for the first time, tracking information in the L1 trigger. Examples of how this tracking information can be used to reduce the L1 trigger rates are presented.}

\FullConference{The 39th International Conference on High Energy Physics (ICHEP2018)\\
		4-11 July, 2018\\
		Seoul, Korea}

\begin{document}

\setstretch{0.99}

The High-Luminosity LHC (HL-LHC) upgrade will allow us to provide high-precision measurements of the standard model (SM), as well as searches for new physics beyond the SM. It is crucial to prepare an upgrade of CMS detector in order to operate efficiently up to integrated luminosities of 3000 fb$^{-1}$ and to fully profit from the HL-LHC capabilities despite the harsh pileup environment of 200 proton-proton interactions per LHC bunch crossing.

The CMS trigger system currently consists of two levels such as the Level 1 (L1) trigger and the High-Level trigger (HLT). The L1 trigger is implemented in hardware processors that receive data from calorimeter and muon systems, generating a trigger signal within 3 $\mu$s, with a maximum rate of 100 kHz \cite{TDR:017}. The main goal of the Phase-2 upgrade of the CMS L1 trigger is to include the tracking information and high-granularity calorimeter within larger trigger latency of 12.5 $\mu$s with a maximum rate of 750 kHz. The upgraded CMS L1 trigger will be a new key element to maintain trigger thresholds as low as possible for high selection capability for events from low-mass physics processes at electroweak mass scales.

Updates to the core trigger algorithms to match track-trigger information with standalone calorimeter or muon trigger objects are presented. First, displaced muons using a track-match Veto have been studied. Muons with a large displacement are not available to trigger at any $p_{\mathrm{T}}$ cut since the current standalone muon $p_{\mathrm{T}}$ assignment applies a beam spot constraint. The preliminary algorithm for the displaced muon shows to drop the beam-spot constraint, but require precision measurements of the muon direction in at least two stations to measure momentum \cite{TDR:017}.

Following the upgrade of detector electronics for the barrel electromagnetic (EM) calorimeters (ECAL), energy measurements with a granularity of (0.0175, 0.0175) in ($\eta$, $\phi$) are feasible by the digitized response of every crystal of the barrel ECAL. The granularity for the upgraded barrel calorimeters is 25 times higher than the input to the Phase-1 trigger comprising trigger towers which had a granularity of (0.0875, 0.0875). The much finer granularity will provide an improvement in position resolution of the EM trigger algorithms which is critical for calorimeter isolation \cite{TDR:017}.

Algorithms of the identification of $\tau$ leptons by the L1 trigger has been updated using $<$PU$>$ = 200 with a Phase-2 simulation of the detector and the corresponding trigger primitives \cite{TDR:017}. A trigger rate of 50 kHz can be achieved at $<$PU$>$ = 200 with $E_{\mathrm{T}}$ $>$ 40 and 25 GeV for algorithms of Phase-2 L1 tracks matched to Phase-1 L1 Taus and a L1 track-based isolation requirement is applied, respectively.

The missing transverse momentum and the scalar summed $p_{\mathrm{T}}$ over all jets in an event has been studied using the CMS particle flow (PF) and the pileup per particle identification (PUPPI) algorithms \cite{TDR:017}. Further studies for jet substructure for heavy-particle tagging, and lepton isolation will be conducted in progress.

The study on the L1 b-tagging trigger based on both inner pixel and outer tracker information is ongoing using the latest Phase-2 CMS tracker geometry. An improvement of the trigger performance for the L1 b-tagging is expected of the high granularity of pixel layers in the inner tracker and tilted barrel geometry in the outer tracker \cite{L1PixTrk}.

Adding tracking information to the L1 trigger will greatly expand the physics potential by the sophisticated algorithms with high selection power in the hardware of the L1 Trigger. Such algorithms are expected to significantly increase the overall CMS trigger performance at HL-LHC.

\end{document}